\documentclass[conference]{IEEEtran}
\IEEEoverridecommandlockouts
\usepackage{cite}
\usepackage{amsmath,amssymb,amsfonts}
\usepackage{algorithmic}
\usepackage{graphicx}
\usepackage{textcomp}
\usepackage{xcolor}
\def\BibTeX{{\rm B\kern-.05em{\sc i\kern-.025em b}\kern-.08em
    T\kern-.1667em\lower.7ex\hbox{E}\kern-.125emX}}
\begin{document}

\title{Modelling the Performance of High Capacity Access Networks for the Benefit of End-Users and Public Policies\\

\thanks{This work was partially supported by the European Union under the Italian National Recovery and Resilience Plan (NRRP) of NextGenerationEU, partnership on “Telecommunications of the Future” (PE00000001 - program “RESTART”).}
}

\author{\IEEEauthorblockN{Antonio Capone}
\IEEEauthorblockA{\textit{DEIB} \\
\textit{Politecnico di Milano}\\
Milan, Italy \\
antonio.capone@polimi.it}
\and
\IEEEauthorblockN{Maurizio Decina}
\IEEEauthorblockA{\textit{DEIB} \\
\textit{Politecnico di Milano}\\
Milan, Italy \\
maurizio.decina@polimi.it}
\and
\IEEEauthorblockN{Aldo Milan}
\IEEEauthorblockA{\textit{Autorità per le garanzie} \\
\textit{nelle comunicazioni (AGCOM)}\\
Rome, Italy \\
a.milan@agcom.it}
\and
\IEEEauthorblockN{Marco Petracca}
\IEEEauthorblockA{\textit{Autorità per le garanzie} \\
\textit{nelle comunicazioni (AGCOM)}\\
Rome, Italy \\
m.petracca@agcom.it}
}

\maketitle

\begin{abstract}
This paper deals with the challenge of modeling the performance of planned ultrabroadband access networks while maintaining technological neutrality and accuracy in measurable quality. We highlight the importance of such modeling also for addressing public funding policies compared to models mainly based on the maximum nominal speed of the access networks, taking also into account the widespread use of measurement tools like "speed test" that have influenced the perceived quality by end-users. We present a performance modelling approach based on the extension of well-known traffic models that accurately characterizes the performance of broadband access networks. We also show how the presented model has been validated with data from two network operators and has been applied to address the recent Italian public interventions for the development of ultrabroadband access networks in market failure areas. 
\end{abstract}

\begin{IEEEkeywords}
Broadband, Access Networks, Speed Test, Model.
\end{IEEEkeywords}

\section{Introduction}
\label{section1}

In the digital economy era, the availability of a high-quality and secure digital infrastructure is essential to support the development of innovative applications and services to the benefit of all  citizens, companies, and public administrations. Indeed, the correlation between the level of diffusion of broadband internet access and economic development has been widely analyzed in the literature\cite{Fal21}.

Based on this concept several international organizations and countries around the world have defined guidelines and initiatives to ensure the development and technology update of the digital infrastructure (see e.g. the ITU initiative Partner2Connect \cite{itu}).
At European level, the objectives set out in the European Electronic Communications Code (EECC) \cite{EECC}  aim at ensuring connectivity and widespread availability in the European Union (EU) of Very High Capacity Networks (VHCN), as identified by the criteria defined by the Body of European Regulators for Electronic Communications (BEREC) \cite{BerecGLVHCN}. In order to meet these needs, the Digital Decade Policy Programme (DDPP) \cite{DDPP} defines concrete targets to be achieved in the Union by 2030. In particular, as regards connectivity the digital target of the DDPP envisages that “\textit{all end users at a fixed location are covered by a gigabit network up to the network termination point, and all populated areas are covered by next-generation wireless high-speed networks with performance at least equivalent to that of 5G, in accordance with the principle of technological neutrality}”. 
%Even the recent proposal by the European Commission for a Gigabit Infrastructure Act \cite{GIA} aims to contribute to the cost-efficient and timely roll-out of VHCN necessary to meet the EU’s increased connectivity needs, consistently with funding initiatives to support the development of ultrabroadband networks in remote or, more in general, less well-served areas. The recently adopted Guidelines on State aid for broadband networks \cite{SAGL} also contribute to speed up the deployment of VHCN in market failure areas by clarifying when public support could be considered in line with competition rules.

In this framework, in order to estimate the quality of connectivity perceived by customers when using applications as well as to properly address any public policies for digital infrastructure development, it is necessary to consider the actual performance of ultrabroadband access networks rather than just the nominal maximum speed offered by operators. This fundamental aspect is highlighted also by BEREC in its Guidelines on VHCN, which stress the need of considering the capability of the networks, both fixed and wireless, to deliver certain performance “\textit{under usual peak-time conditions}”. % (e.g. in terms of downlink and uplink data rate – 1.000 and 200 Mbps for fixed connections, 150/50 Mbps for mobile connections).

In the past years, a series of commercial "network speed test" applications have been developed and made available to the market, gaining increasing popularity \cite{ookla,fast}. Despite numerous limitations, such tools have become an user-friendly evaluation metric for consumers to assess whether the subscribed fixed/mobile network meets their quality expectations and the promises of operators commercial offers \cite{Fea20}. The same fixed/mobile operators have started to develop their own dedicated apps to allow end-users to perform speed tests and to collect information by end-users on the perceived quality of their services through instant surveys. Besides that, in order to ensure a high level of transparency and user protection, numerous National Regulatory Authorities (NRA) for telecommunications adopted similar tools and regularly conduct extensive measurement campaigns aimed at providing end-users with comparison of the actual performance delivered by the fixed and mobile access networks of different operators.

The use of the aforementioned measurement tools is certainly helpful to provide a detailed picture of the performance of networks that have already been deployed and activated. However, such tools are basically not usable in the planning phase of infrastructure development where a modelling approach is necessary for guiding strategies and decisions on possible evolution at both private and public levels. For these reasons, a valid model is needed to characterize the expected performance of VHCN in the planning phase taking into account as real as possible traffic conditions, as most technologies rely on resource sharing among end-users and the network capacity actually usable by each end-user depends on the level of traffic. Moreover, such a model should be flexible enough to be validly applicable for different kind of fixed and wireless communication technologies that can be implemented to achieve the VHCN performance, in compliance with the principle of technological neutrality. 

In this article, we address the problem of how to model the performance of VHCN while remaining neutral with respect to the technology used and maintaining a good level of accuracy with respect to ex-post measures of the quality of service. We present a performance model based on the adaptation of established traffic models and its validation with data from two network operators. Furthermore, we present the case of Italian public policies for the development of VHCN implemented in the National Recovery and Resilience Plan (NRRP), where the proposed model was adopted in the mapping exercise carried out according to the State aid rules as well as in the tender procedures to assign public fundings. In Section \ref{section2}, we review the approaches used in characterizing access networks and the definitions used by regulatory bodies (with particular reference to the European case). In Section \ref{section3}, we present the proposed modelling approach derived from the classical processor sharing model adapted to access network, while in Section \ref{section4} we discuss its characteristics and some validation results. In Section \ref{section5}, we present the case of Italian plans for the development of broadband access, and finally, in Section \ref{section6}, we draw some concluding remarks.

\section{Definitions and measurement approaches commonly adopted}
\label{section2}
Measuring the performance of access networks can have different objectives, and the most suitable modeling and measurement tools depend on them.

\begin{figure}[hbt]
\begin{center}%
\includegraphics[width=0.90 \columnwidth]{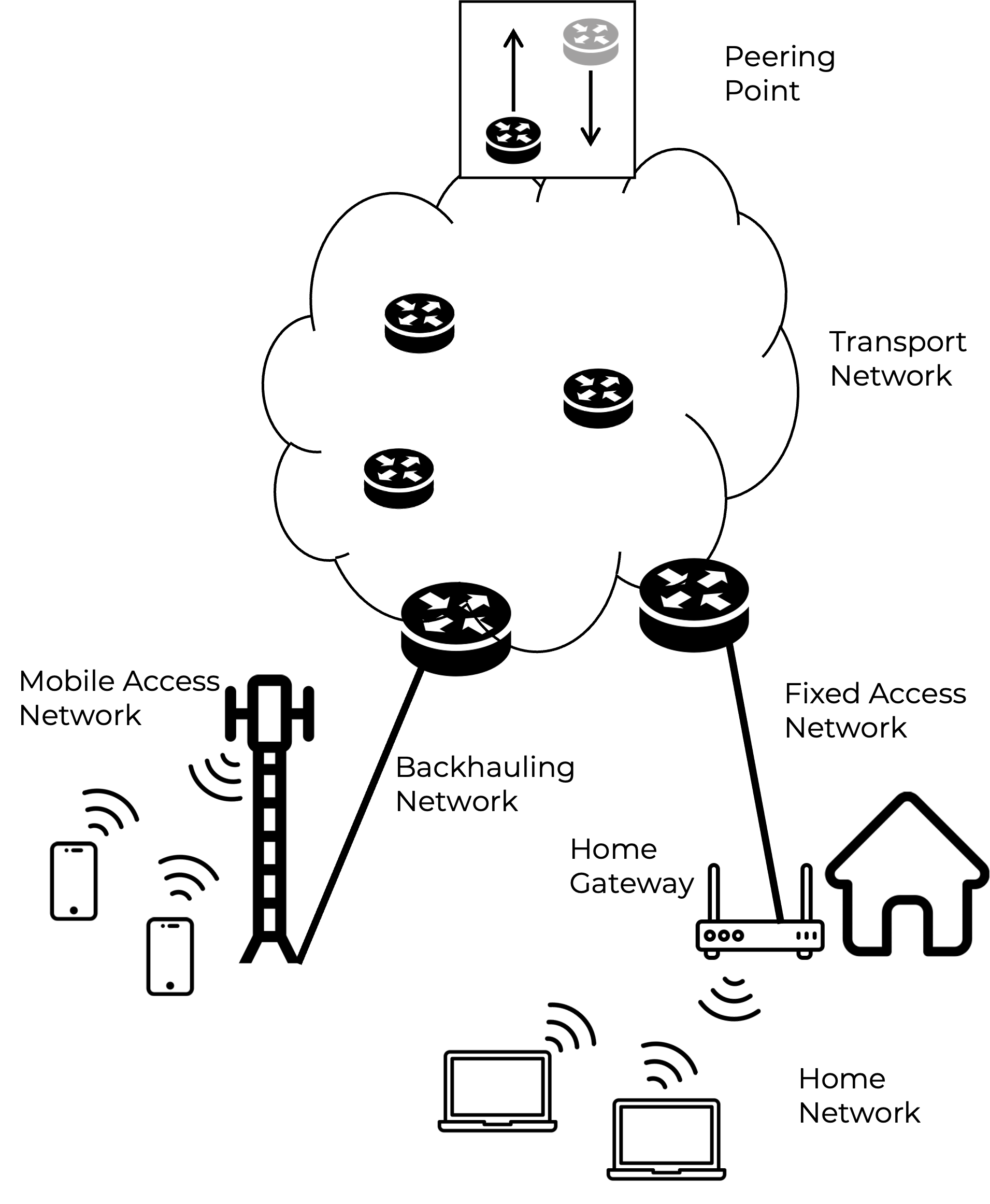}
\end{center}
\caption{Internet Access Provider typical network architecture.}
\label{AccessNetwork}
\end{figure}

A first objective may be to provide end-users with quantitative indicators directly linked to the quality of experience in using major applications. In this case, the quality of experience depends on the end-to-end connection between the user terminal and the servers of the applications used and is potentially influenced by a number of elements along the path, not all of which are under the control of the Internet access service provider. In a typical Internet access network architecture, as shown in Figure \ref{AccessNetwork}, the connection speed depends on the system bottleneck, which is not always represented by the access network, especially in recent years with the greater diffusion of ultra-broadband access \cite{Fea20}. The limitation may be external to the operator's network and depend on the Content Distribution Network (CDN) used by the application provider. This explains why some bandwidth-intensive service providers have started using their own measurement applications \cite{fast} to give users an estimate of the available capacity to their servers (unfortunately with possible discriminations between access operators). The limitation may be in the user's Home Network due to the capacity limits of the gateway or the domestic distribution technology used (typically Wi-Fi). This aspect can be excluded from the measurement, for example, by adopting speed testing applications installed directly in the Home Gateway \cite{Sun11}. In the case of mobile access networks several approaches have been proposed \cite{Man18}, but with the increasing capacity of technologies such as 5G, the limitation may reside in the backhauling network, which for many operators is still partially based on point-to-point radio links with limited capacity. Many of the tools currently used by end-users are unable to discriminate which element is limiting the connection speed, and one of the challenges is certainly to develop new tools that provide users with an undistorted view of the quality of their Internet access.

A second objective may be that of NRA to provide end-users with a fair comparison of the quality of Internet access networks among national operators offering commercial services in a competitive environment. In order to achieve this goal, it is necessary to ensure that the measurements are not influenced by external factors outside the operator's network and that they are representative of the connection between the end-user and the point of connection with other operators (Peering Point). Many of the NRA that carry out measurement campaigns adopt various strategies to ensure a fair comparison between operators. Typically, these operators collaborate to strategically place measurement servers and select measurement protocols that are not influenced by external parameters \cite{ofcom, fcc, agcom}.

Finally, a third objective could be to address public funding policies for the development of Internet access infrastructure with the aim of stimulating economic growth and ensuring the availability of \textit{future-proof} networks able to provide end-users with adequate quality of services to satisfy their current and evolving needs, even in peripheral areas where market failure occurs. Until recently, development objectives for access infrastructure have often referred to the maximum nominal speed provided by network technology (achievable under ideal conditions with a single user connected) and often ignored aspects of traffic due to resource sharing. Some models have attempted to introduce the resource sharing component with "contemporaneity factors" aimed at calculating the average number of simultaneously connected users. In 2020, BEREC has published its guidelines on VHCN, proposing an approach that directly refers to the speed actually experienced between the user terminal and the peering point under peak traffic conditions, somewhat in line with the measurement mechanisms commonly used by speed test applications and NRA measurement protocols \cite{BerecGLVHCN}.

Guidelines such as those from BEREC can provide the basis for the implementation of strategic policies such as the DDPP, which defines connectivity goals of the EU in the medium to long term, as well as national policies for public investments in market failure areas where the achievement of these goals cannot be guaranteed by private investments. In this context, the use of measurement campaigns can provide a snapshot of the 'status quo' (i.e. the existing networks), while a model is necessary to plan intervention measures for new ultrabroadband access infrastructure. Such a model, although with necessary approximations, should be able to capture the fundamental elements (including network capacity and traffic) that determine 'ex-post' the measurable speed.

\section{Proposed modelling approach}
\label{section3}

End-to-end measurement techniques can measure network performance without explicitly identifying the bottleneck link. However, in a model designed to guide the development of access networks, we need to focus solely on the shared access link, disregarding other possible bottlenecks. If limitations to the end-to-end performance arise from other elements in the network segment between the user terminal and the peering point, specific actions should be taken to remove them. These limitations may be due to insufficient capacity of other links in the network, such as the backhauling links of mobile access, or to operator traffic management strategies, such as traffic shaping techniques implemented in network gateways.  

In order to define a model able to follow the behavior of speed test tools when the bottleneck is in the access network, it is necessary to consider two aspects: i) how measurement protocols typically work and ii) how network manages available access resources. Measurement protocols generally perform long file transfers using TCP connections and then estimate the speed as the file size $x$ divided by the transfer time $d(x)$. Some corrections to measured values are adopted to compensate for the initial slow start phase of the TCP connections. As for the access resource management, transmission scheduling algorithms generally utilize "fair scheduling" policies that equally divide available resources among active flows. For access technologies with multiple Modulation and Coding Schemes (MCS) that adjust the rate according to current channel conditions, such as wireless technologies, scheduling policies often rely on "Proportional Fair" scheduling. This ensures that the average rate per active flow is proportional to the rate of the MCS being used.

The most appropriate traffic model able to describe scheduling policies that fairly assign all available resources among active flows, is the well known "processor sharing" model \cite{kleinrock}. This model is also in line with the behavior of so called elastic flows, like TCP connections that fairly share all the available capacity of the bottleneck link.

\subsubsection{Fair sharing}
\begin{figure}[hbt]
\begin{center}%
\includegraphics[width=0.80 \columnwidth]{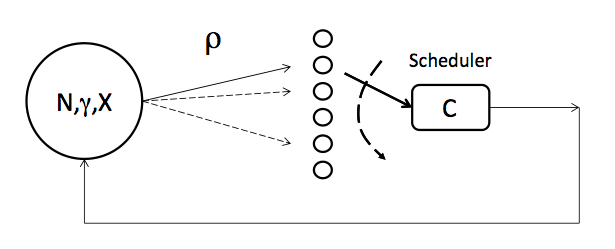}
\end{center}
\caption{Simple model with fixed number of users and constant capacity.}
\label{queue}
\end{figure}

Let's consider a shared channel model among multiple users (as shown in Figure \ref{queue}), such as the radio channel of a cell or the optical connection of a GPON, with the following characteristics:
\begin{itemize}
\item $N$ users, each of whom generates a service request (data transfer) of $X$ bits (random variable) at a rate of $\gamma$ (Poisson process) and does not generate any other traffic until the previous request has been served;
\item a channel with a capacity of $C$ bit/s shared among the users through a resource scheduler;
\item service requests in the system are served by equally dividing the channel capacity between them (processor sharing).
\end{itemize}

The model describes the behavior of a shared access channel under general conditions. It belongs to  the class of "symmetric" queues where, whatever is the distribution of $X$, the occupancy and crossing time of the processor (transmission time) are the same, and equal to the easy-to-solve case where $X$ is exponentially distributed.

Therefore the model is equivalent to an M/M/1 queue with finite population $N$. The performance indicators of this model can be easily expressed in closed form, and they are not much different from that with infinite population (Poisson model). For simplicity, we consider in the following the formulas for the infinite population case. 

The average transmission time is given by: 
\begin{equation}
D  =\frac{m_X/C}{1-\rho}\label{eq: 73004aa}
\end{equation}
where $m_X$ is the average of $X$, $\rho=\frac{\lambda m_X}{C}$ and $\lambda=N\gamma$.

A property of symmetric queues is that the same formula holds for a given service $X=x$: 
\begin{equation}
d(x)=\frac{x/C}{1-\rho} \label{eq: busy8}
\end{equation}

In our case $d(x)$ is the time required to transfer $x$ bits and $C$ is the access channel capacity. The average transfer rate (per-user throughput) is by definition $v=x/d(x)$ and then:
\begin{equation}
v=(1-\rho)C \label{eq: busy88}
\end{equation}
Note that the average rate $v$ is different from the instantaneous rate $C/n$ since the number $n$ varies during the transfer period.
\subsubsection{Class-based Fair sharing}
Let us consider now the case of fair sharing with different classes of users characterized by different frequencies $\lambda_i$ and different requests $X_i$ with mean $m_{i}$. Also this case falls within the symmetric queue models, and the same results of previous case also holds here with: 
\begin{equation}
\rho=\frac{\lambda m_X}{C}=\sum_i  \frac{\lambda_i m_i}{C}=\sum_i \rho_i.  \label{eq: busya80}
\end{equation}

The average service (transmission) time is the same in eq. (\ref{eq: 73004aa}), while the average service time for class $i$ is:
\begin{equation}
D_i  =\frac{m_i/C}{1-\rho}.
\label{eq: 7ab}
\end{equation}

The average service time for a user of class $i$  requesting the trasfer of $x_i$ bits is:
\begin{equation}
d(x_i)  =\frac{x_i/C}{1-\rho}\label{eq: 7ac}
\end{equation}
and the average transfer rate is: 
\begin{equation}
v_i=(1-\rho)C
\end{equation}

\subsubsection{Proportional fair sharing}
Let us now model the case in which also the channel capacity depends on the user class, i.e. the case of multiple MCSs with rate selected based on channel quality. Let us assume that, if there is a single user in the system, the processor provides a capacity $C_i \le C$, (reference channel rate
for class $i$ based on the corresponding MCS), and that the instantaneous rate $c_i$ of a user in class $i$ in the presence of other users of an arbitrary class $j$ is such that:
\begin{equation}
\frac{c_i}{C_i}= \frac{c_j}{C_j} = \alpha . 
\end{equation}

The total capacity can be written in two different ways. We obviously have:
\begin{equation}
C'=\sum_{j=1}^n c_j = \sum_{j=1}^n \alpha C_j
\end{equation}
where $n$ is the number of users (active flows) in the processor (scheduler). But, observing that in the unit time class $i$ transmit $C_i$ bits, we also have:
\begin{equation}
C' = \frac{\displaystyle \sum_{j=1}^n C_j}{n}
\end{equation}
which provides
\begin{equation}
\alpha= \frac{c_i}{C_i}= \frac{1}{n}.  \label{eq: x3}
\end{equation}
The capacity $C' \le C$ then depends on the type of users, while $\alpha$ only from their number.

This type of queue is no longer a symmetric queue. In the case of the proportional fair scheduling, the time $\Delta T$ for processing (transmitting)  $\Delta x$ bits is:

\begin{equation}
\Delta T_i=\frac{\Delta x}{ c_i}= \frac{\Delta x }{\alpha_i C_i}= \frac{\Delta x \ n}{C_i}.
\label{eq: b3}
\end{equation}
while in the class-based case it is: 
\begin{equation}
\Delta T_i=\frac{\Delta x}{c_i}= \frac{\Delta x \ n}{C}. \label{eq: b3bis}
\end{equation}

Comparing with the proportional fair we see that we can write:
\begin{equation}
\Delta T_i=\frac{\Delta x \ n}{C_i}=\frac{\Delta x' \ n}{C}. \label{eq: b3ter}
\end{equation}
where
\[\Delta x'=  \Delta x \frac{C}{C_i} = \Delta x  \nu_i \]

which means that the \emph{class-based fair sharing} case has the same transmission time of the 
\emph{proportional fair} if, we increase the amount of bits to be transmitted by $ \nu= C/C_i$.

We then need to consider new service requests with average 
\begin{equation}
m'_X= \sum_i \frac{\lambda_i}{\lambda} \nu_i m_X, \label{eq: b4}
\end{equation}

So we have
\begin{equation*}
\rho= \rho'=\frac{\lambda m'_X}{C}=\sum_i \frac{\lambda_i \nu_i m_X}{C}=  
\end{equation*}
\begin{equation}
=\sum_i \frac{\lambda_i
m_X}{C_i}=\sum_i \frac{\lambda_i }{\lambda}\frac{\lambda m_X}{C_i}=\sum_i \frac{\lambda_i
}{\lambda}\rho_i \label{eq: busya82}
\end{equation}
being $\rho$ the fraction of time in which the processor is active, and $\rho_i$ the fraction of time spent to serve class $i$.

The service (transmission) time for a given service request $x$ is then:
\begin{equation}
d_i(x)=\frac{x' /C}{1-\rho'} =\frac{x /C_i}{1-\rho} \label{eq: 7aca}
\end{equation}

The equation:
\begin{equation}
\rho=1-\frac{x/C_i}{d_i(x)}= 1-\frac{v_i}{C_i}. \label{eq: busy19}
\end{equation}
allows to measure $\rho$ given the average service rate $v$ and the channel rate $C_i$ for class $i$ .

While the transfer rate per-user for class $i$ is the same of the class-based fair sharing case:
\begin{equation}
v_i=(1-\rho)C_i. \label{eq: busy119}
\end{equation}

\section{Validation and discussion}
\label{section4}

The model allows to estimate, independently from the access technology, the average connection speed per user $v_i$, given the nominal channel rate $C_i$ and average traffic load $\rho$, under the general assumption that the resource scheduling if proportional fair (with the simple fair scheduling being a special case). 

The applicability of the model for the case of fiber access based on Passive Optical Network (PON) technologies is rather straightforward since in most scenarios the modulation scheme is equal for all users and fairness in scheduling transmissions is in general guaranteed. Much more interesting is the case of wireless access, either fixed or mobile, where the use of multiple modulation and coding schemes, the scheduling policies adopted and the impact of interference may limit the accuracy of the model. 

We have been able to conduct an experimental validation of the estimates provided by the traffic model for wireless mobile networks with the support of two  operators as part of the collaboration with the Italian Minister for Technological Innovation and Digital Transition for the definition of the national public plans for ultrabroadband access described in the next section. The results shown here are just examples of a larger validation campaign that we have been able to conduct considering:
\begin{itemize}
\item test campaigns carried out with different "speed test" protocols and by different entities;
\item a broad set of scenarios with cities of different sizes in some European countries;
\item different manufactures of radio equipment, verifying that the behavior of radio schedulers is always reasonably similar to that of proportional fair;
\item in different carrier aggregation scenarios, showing that the accuracy of the single carrier model is also preserved in the case of aggregated carriers.
\end{itemize}

\begin{figure}[bt]
\begin{center}%
\includegraphics[width=0.90 \columnwidth]{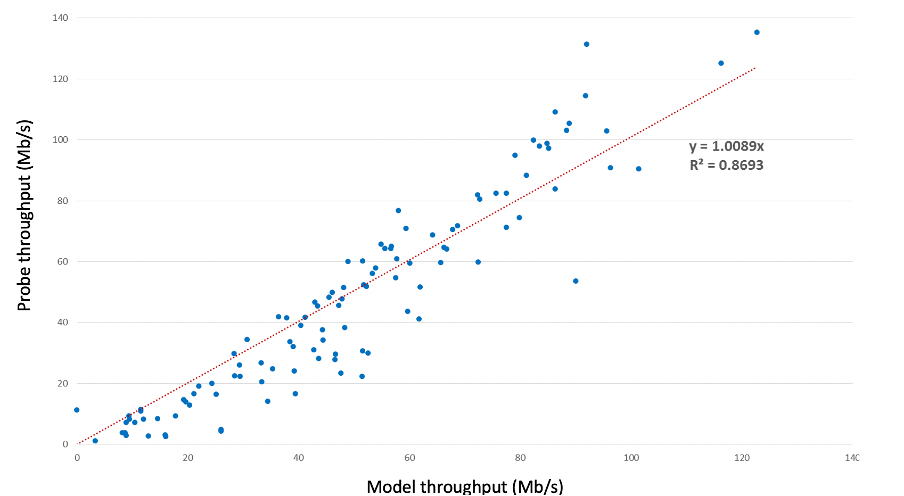}
\end{center}
\caption{Validation of the model with speed tests in mid-size city.}
\label{validation1}
\end{figure}

\begin{figure}[bt]
\begin{center}%
\includegraphics[width=0.90 \columnwidth]{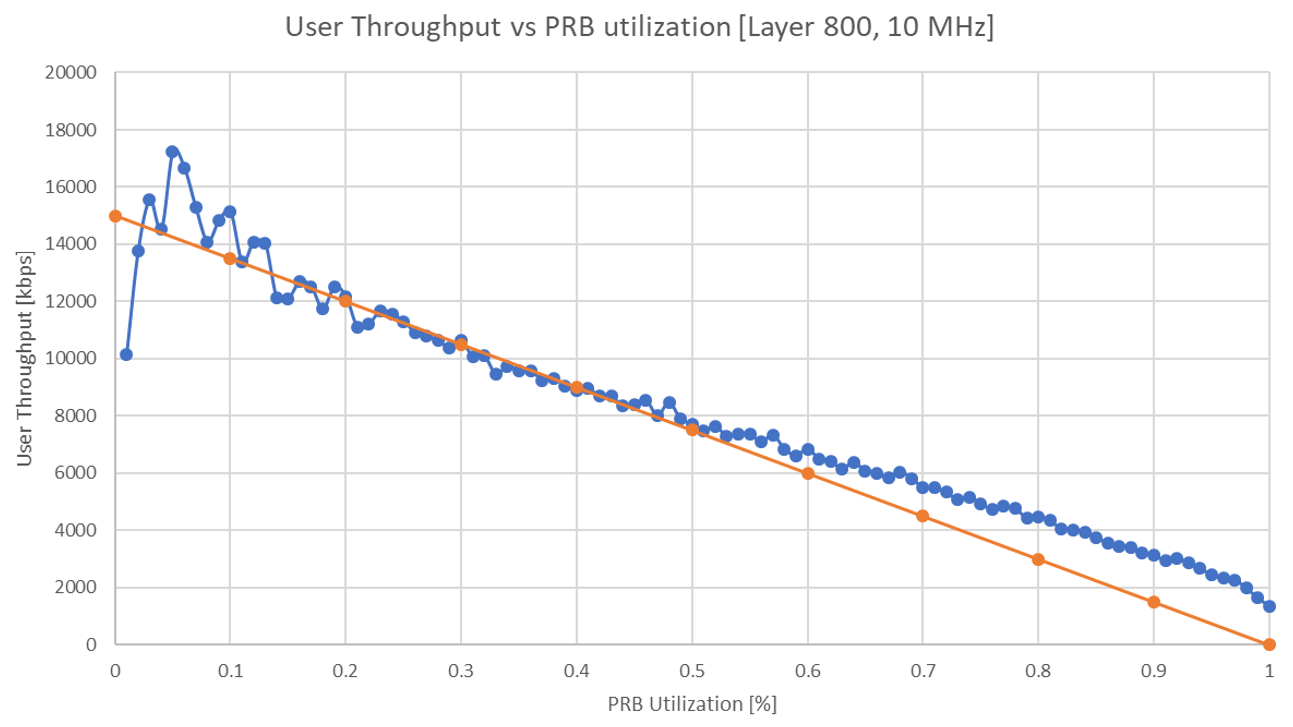}
\end{center}
\caption{Validation of the model with network-wide throughput indicators extracted from LTE network at 800 MHz band.}
\label{validation2}
\end{figure}

\begin{figure}[bt]
\begin{center}%
\includegraphics[width=0.90 \columnwidth]{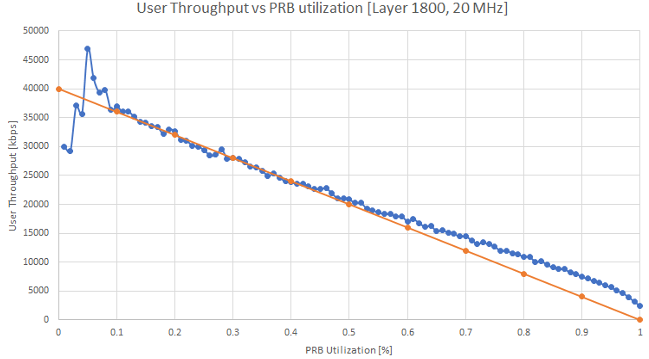}
\end{center}
\caption{Validation of the model with network-wide throughput indicators extracted from LTE network at 1800 MHz band.}
\label{validation3}
\end{figure}

We describe, for example, the case of a measurement campaign in a medium-sized Italian city, in the case of aggregation of two LTE carriers. Figure \ref{validation1} shows a scatter-plot in which each point indicates a speed test, while the speed value estimated by the model is shown in the $x$ axis and the speed value measured by the probe is shown in the $y$ axis. The model is evaluated by detecting the Modulation and Coding Scheme (MCS) reported by the probe at the time of the measurement to calculate $C_i$, and the average occupancy value of PRBs (Physical Resource Blocks) from the statistical counters of the radio base station on the date and time the measurement to calculate $\rho$. Being the measurement carried out in Carrier Aggregation (dual carrier in particular), the estimation of the per user rate was calculated for both the primary and secondary carriers. For the latter, the MCS was assumed equal to that of the primary because the information about the one actually used is not reported in the speed test record. The accuracy of the model is rather good considering that points plotted are individual tests and that some fluctuations of the measured values in this kind of experiments are unavoidable.

A second validation example has been provided by another Italian operator which has compared the values provided by the model with the aggregated per user-throughput indicators provided by the whole national LTE network (with more than 20.000 base stations). The plots are shown in Figure \ref{validation2} for the 800 MHz cells and in Figure \ref{validation3} for the 1800 MHz cells. Each figure shows the values measured in the network (blue dots and line) and the values predicted by the model (orange dots and line) versus the different values of $\rho$ averaging for each value of $\rho$ the indicators reported in different cells of the national network. As we can observe the model is accurate for a large mid range of $\rho$ values. At low values of $\rho$ the values measured in the network tend to be more dispersed above and below the value estimated by the model probably due to the limited samples available. At very high values of $\rho$ the rates measured in the network tend to be higher than those predicted by the model probably because of admission control policies that select a subset of users in scenarios of high network congestion.

\section{The experience of the Italian public funding plans for ultrabroadband access}
\label{section5}

The modelling approach proposed above has been recently applied by the Italian Government for the public plans aimed at financing both fixed and mobile ultrabroadband access networks in market failure areas under the NRRP\cite{SBUL}.

In particular, the model has been adopted for the following two public plans: i) the "Italia a 1 Giga" Plan for the development of fixed access infrastructure able to provide all the identified households in market failure areas with 1 Gbps in download and 200 Mbps in upload under usual peak time conditions (according to the BEREC guidelines on VHCN); ii) the "Italia 5G" Plan for realizing in market failure areas optical fibre backhauling and new base stations able to provide 150 Mbps in download and 30 Mbps in upload under usual peak traffic conditions (in line with BEREC indications on VHCN). The modelling approach has been discussed with stakeholders in public consultations and some operators have also collaborated for its validation (as mentioned in previous Section).

For both the Plans, a detailed mapping of the ultrabroadband infrastructure has been carried out to identify eligible areas for State aid intervention, i.e. those areas where the achievement of specific performance thresholds can not be ensured by private investments. In particular, all the involved operators have provided snapshots of their infrastructure both existing and expected in the next years until 2026 (deadline set by the NRRP) based on their business plans. 

For fixed networks, all households with a civic address have been mapped and the estimation of the per-user rate in peak traffic conditions has been calculated using the model. In particular, the nominal channel rate per user has been used for estimating $C_i$ according to the specific technology used and the traffic per user has been used for estimating $\rho$ based on the set of users sharing the same access resources. As for the snapshot of the existing infrastructure, the traffic has been estimated considering the available measurements, while for the next years a traffic growth rate based on the traffic data collected by AGCOM have been estimated.

For the mapping related to the "Italia 5G" Plan, radio propagation tools have been used for the simulation of the radio coverage and channel quality over the entire national territory with a resolution of 100 $\times$ 100 metres (pixel). As for the traffic, considering the high variability in the pixel due to the typical characteristics of mobile networks, the calculation of $\rho$ has been done by operators based on their traffic estimates per cell (accounting for different technologies and frequency layers available).

For both Plans, the described modelling approach has been adopted in the tender procedures for assigning the public funding for realizing the infrastructure.

\section{Conclusion}
\label{section6}

In this paper we have discussed the issues associated with the modelling of the performance of planned ultrabroadband access networks. In particular, we have shown that in addition to the measurement approaches for existing infrastructure commonly used by end-users, operators and NRA, a modelling approach for the definition of public policies in the development of the ultrabroadband access infrastructure is necessary. We have presented the proposed modelling approach based on an extension of the well known queuing systems. This model was recently applied in the Italian funding plans for the development of VHCN supported by the EU under the NRRP. We have also shown how the proposed model has been validated with the support of some Italian operators.

\section*{Acknowledgments}
The authors wish to thank prof. Flaminio Borgonovo for the fruitful discussions on the traffic model and its applicability to the dimensioning of access networks. 
The authors also wish to thank the Italian operators that have contributed in the validation of the model. The views expressed in this paper are those of the authors and they
should not be assumed to represent Agcom position.

\end{document}